\documentclass[a4paper,11pt]{article}
\usepackage{pos}
\usepackage[binary-units=true]{siunitx}  

\title{The upgraded Data Acquisition System of the H.E.S.S. telescope array}
 \ShortTitle{The upgraded H.E.S.S. DAQ}

\author*[a]{Sylvia J. Zhu}
\author[a]{Tim Lukas Holch}
\author[a]{Thomas Murach}
\author[a]{Stefan Ohm}
\author[a]{Matthias F\"u{\ss}ling}
\author[b]{Mathieu de Naurois}
\author[a]{Fabian Krack}
\author[a]{Klemens Mosshammer}
\author[a]{Rico Lindemann}

\affiliation[a]{Deutsches Elektronen-Synchrotron (DESY), D-15738 Zeuthen, Germany}
\affiliation[b]{Laboratoire Leprince-Ringuet, CNRS, Institut Polytechnique de Paris, F-91128 Palaiseau, France}

\emailAdd{sylvia.zhu@desy.de}

\abstract{The High Energy Stereoscopic System (H.E.S.S.) is an array of five Imaging Atmospheric Cherenkov Telescopes located in the Khomas Highland of Namibia. H.E.S.S. observes gamma rays above tens of GeV by detecting the Cherenkov light that is produced when Very High Energy gamma rays interact with the Earth’s atmosphere. The H.E.S.S. Data Acquisition System (DAQ) coordinates the nightly telescope operations, ensuring that the various components communicate properly and behave as intended. It also provides the interface between the telescopes and the people on shift who guide the operations. The DAQ comprises both the hardware and software, and since the beginning of H.E.S.S., both elements have been continuously adapted to improve the data-taking capabilities of the array and push the limits of what H.E.S.S. is capable of. Most recently, this includes the upgrade of the entire computing cluster hosting the DAQ software, and the accommodation of a new camera on the large 28m H.E.S.S. telescope. We discuss the performance of the upgraded DAQ and the lessons learned from these activities.}

\FullConference{37$^{\rm{th}}$ International Cosmic Ray Conference (ICRC 2021)\\
		July 12th -- 23rd, 2021\\
		Online -- Berlin, Germany}

\begin{document}
\maketitle

\section{Introduction}

One of the advantages of ground-based observatories is that they can be 
upgraded and adjusted to fit the changing needs of the collaboration that runs 
them. The High Energy Stereoscopic System (H.E.S.S.) has undergone multiple 
major upgrades in its nearly two decades of operations. Originally commissioned 
in 2003 as a set of four 12-m telescopes (CT1-4), the addition of a 28-m 
telescope (CT5) in 2012 greatly extended the array's sensitivity to lower photon 
energies \cite{HESSII_1, HESSII_2}. The CT1-4 cameras were upgraded 
in 2016 and the CT5 camera in 2019, both times to  
replace the aging electronics with modern technologies,
ensuring stable operations and allowing for novel data-taking modes.

A central computing system coordinates the activities of the array.
This brain of the array is the Data Acquisition System (DAQ); it communicates commands
to the array subsystems, and stores the data that is taken. The DAQ comprises both
the software that manages the array operations --- including the Graphical 
User Interface (GUI), with which the shift crew guide the telescope operation 
--- and the hardware that houses it.

The core philosophy behind the DAQ is described in \cite{Balzer+2013} and remains
extremely relevant. However,
by 2018, many hardware components 
had reached or would soon reach their "end-of-life"
status and were no longer supported by their vendors. Aging, outdated components became
difficult to replace; when they \emph{were} replaced by newer equivalents, this
increased the heterogeneity of the cluster thereby increasing maintenance costs.
Additionally, the existing and planned camera upgrades introduced technological requirements
that the old cluster hardware could not meet. 
In order to ensure smooth
operations for at least seven years, the estimated minimal 
remaining lifetime of H.E.S.S. at the time,
it was decided
to undertake a full upgrade of the DAQ hardware; this included the main cluster
as well as supporting devices such as the switches and firewall.

In addition to the hardware, the operating system (OS) deployed on the previous DAQ cluster
was long past the end-of-life date, so that no upgrades and especially security updates
were provided. Therefore it was decided to upgrade the operating system from Fedora Core
12 to CentOS 7, which at the time of writing is supposed to receive updates
until June 2024.

The upgrade was planned with the following key concepts:
\begin{itemize}
\setlength\itemsep{-0.4em}
    \item The system must be sufficiently future-proof, both in terms of hardware and software.
    \item The cluster must have built-in redundancy, with spares that could easily be swapped in.
    \item The long-term maintenance effort should be minimized when possible.
    \item The setup should be easily reproducible.
    \item The hardware and OS should support the needs of the collaboration software.
\end{itemize}


A comprehensive description of the DAQ can be found in \cite{Balzer+2013}. Here,
we focus on the differences in implementation between the previous iteration and the
most recently upgraded DAQ.

\section{Cluster setup}

\begin{figure}
\centering
\includegraphics[width=0.85\textwidth]{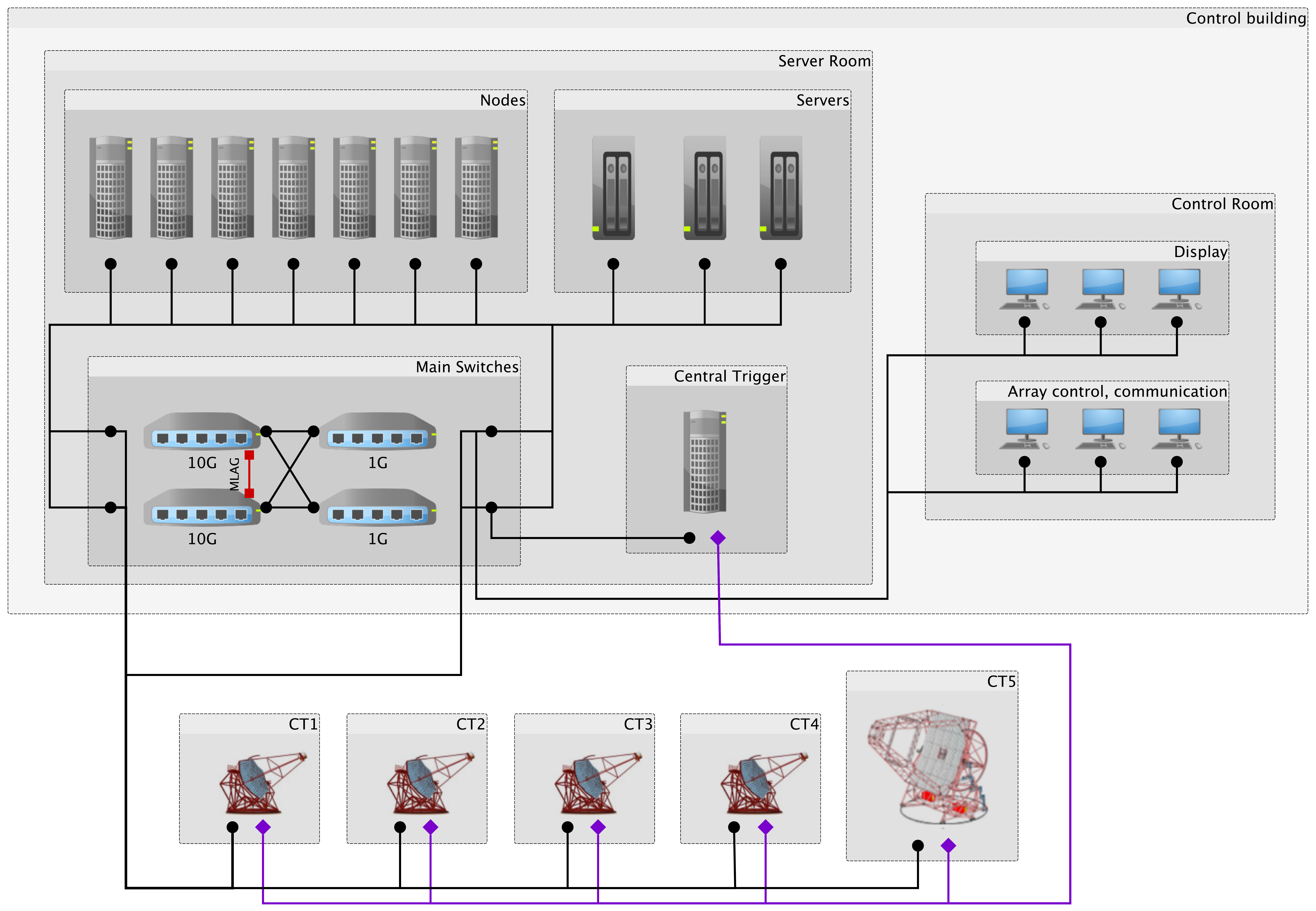}
\caption{Network diagram of the DAQ machines and the data-taking components of the array. Black lines with circle endpoints indicate network ethernet connections. The purple lines with the diamond ends show the direct optical fiber connections between the cameras and the Central Trigger. The red lines with squares show the MLAG setup between the 10Gb/s switches.}
\label{fig:network}
\end{figure}

The core DAQ cluster is composed of seven computing nodes and three storage servers, and
is illustrated in Figure~\ref{fig:network}.
During data taking, the cameras on the telescopes report the detection of air shower
events to the Central Trigger. A candidate event in CT1-4 must be accompanied by
the reporting of a coincident event (within $\approx \SI{80}{\nano\second})$ in one
of the other four telescopes (including CT5), otherwise it is dropped \cite{CentralTrigger}; 
events in CT5 do not have this requirement and are automatically saved \cite{Bolmont+2014}.
The recorded event camera
images are sent to the computing nodes in a round-robin load balancing scheme
\cite{roundrobin}. Each computing node receives chunks of \SI{4}{\second}
of data, which are then transformed into the target data format.
The full set of data for each run (including auxiliary files describing the array
statuses during the run) is then sent to the storage servers, before being transferred
over the internet to clusters in Europe. 

As with the previous iteration of the DAQ \cite{Balzer+2013}, as soon as the new cluster
hardware had been decided upon, a scaled-down 
version of the upgraded DAQ cluster --- a TestDAQ --- was installed at one of 
the home institutions, for studying the key components
of the new cluster. 

\subsection{Core cluster: computing nodes and storage servers}
\label{sec:core}

\begin{table}
    \centering
    \begin{tabular}{c|c|c|c|c|c}
         Type & Quantity & CPU & RAM & Storage & Network\\
         \hline
         \hline
         computing & 7 & Intel Xeon Silver 4114, & 96 GB & 4 $\times$ 2 TB & 2 $\times$ 10 Gb/s\\
         nodes & &  10 $\times$ 2.2 GHz (20 threads) & & & + 2 $\times$ 1 Gb/s\\
         \hline
         storage & 3 & Intel Xeon Silver 4114, & 48 GB & 12 $\times$ 8TB & 2 $\times$ 10 Gb/s\\
         servers & & 10 $\times$ 2.2 GHz (20 threads) & & (RAID6) & + 2 $\times$ 1 Gb/s\\
          \hline
         management & 1 & Intel Xeon Silver 4114, & 48 GB & 2 $\times$ 480 GB & 2 $\times$ 10 Gb/s\\
         machine & & 4 $\times$ 2.6 GHz (8 threads) & & & + 2 $\times$ 1 Gb/s\\
         \hline
         control room & 6 (+2) & Intel Core i3-7100 & 8 GB & 120 GB & 2 $\times$ 1 Gb/s\\
          machines & & 3.9 Hz dual core & & & \\
         \hline
    \end{tabular}
    \caption{Hardware specifications of the machines of the DAQ cluster (Sections~\ref{sec:core} and \ref{sec:controlRoom}).}
    \label{tab:specs}
\end{table}

Compared to the setup described in \cite{Balzer+2013}, the new cluster 
machines reflect the updated needs of the array. New data-taking modes and the upgraded
cameras have greatly increased the rate at which disk space is used; 
currently, approximately 30--40~TB of 
data is saved every month, compared to less than 5~TB per month in the months
preceding the DAQ upgrade.

Until 2018, due to the lack of a fast, reliable, and cost effective internet
connection at the time, the data were stored on magnetic tapes that were manually transported
to Europe. Nowadays, the H.E.S.S. site is on a 100 Mbps plan, allowing
for rapid transfer of the data. However, 
breakdowns in internet communications can occur especially between the continents,
and it is therefore crucial to have enough storage space onsite to maintain a comfortable
buffer. Each of the three storage servers (Table~\ref{tab:specs}) is composed of
twelve 8~TB disks in a RAID6 \cite{RAID6} configuration with one disk reserved as a hot swap,
so that the total amount of available storage space onsite is \SI{216}{\tera\byte}.
In practice, after allowing for space for user homes, database storage, and other needs,
the amount of available storage for camera data is \SI{188}{\tera\byte}.

The computing nodes serve multiple purposes. Six of the seven nodes receive the 
camera data and process the data so that it can be 
stored in a standard format. As they process the data, they simultaneously 
run real-time analyses \cite{RTA} 
on it as it is being taken, to provide low-latency results at somewhat reduced sensitivity 
compared to a full offline analysis. A set of medium-latency, moderate sensitivity 
analyses are also run on hours-timescales;
during the night, they are only run on the seventh node (which is not used in
data taking), while during the day they are run on all but one of the computing nodes 
to increase computational resources. Finally, the computing nodes house the DAQ
software, and the seventh node serves as the login node to the internet and is 
accessible from outside via a hardware firewall.

A single storage volume, comprising dedicated fractions of the storage capacity of
each of the three servers, is formed by means of a distributed GlusterFS file system 
\cite{GlusterFS}. All computing nodes mount this single GlusterFS volume and can thereby
easily and transparently read from and write to the data storage server compound.


In addition, there is a network management machine for the firewall and switch
configuration (described in the sections~\ref{sec:internalNetwork} and \ref{sec:external}).

\subsection{Control room}
\label{sec:controlRoom}

The telescope array is operated from a main control room which hosts a number of general purpose Linux machines that provide interfaces to the DAQ GUIs and monitoring displays.
Over the years of operation, the control room set up was gradually extended as the experiment evolved, which resulted in a heterogeneous hardware setup with task-specific machines and thus high maintenance demands.
The control room was therefore restructured to a homogeneous hardware setup in the first step of the cluster upgrade.
In its current state, the array control interface with all its monitoring displays and GUIs requires a total display area of around twelve 24" full-HD screens (1080$\times$1920 pixels) in order to be well readable for the shift crew.

The upgraded control room makes use of four general purpose Linux machines with 3 HDMI ports each. A fifth machine is provided as a hot swap and can take over any task of the other four machines by simply changing display host entries in the main H.E.S.S. DAQ database, while a sixth primarily provides communication channels to the rest of the collaboration. Further redundancy is achieved by having two additional machines set up as cold spares.


\subsection{Internal network}
\label{sec:internalNetwork}

The communication within the internal network is handled 
by a set of two 10~Gb/s switches (Arista 7050TX-64-R) and three 1~Gb/s switches (Arista 7010T-48-R). One of the latter switches serves as a backup. The two 10-Gb/s switches form a Multi-chassis
Link Aggregation Group (MLAG) for redundancy and optimised transfer rates. The switches are accessible via
the management machine and a dedicated serial connection, set up using a 
terminal server (Perle IOLAN STS) on a dedicated network, exclusively. In the previous DAQ system, the internal network
was supported by only 1~Gb/s switches, without any built-in redundancy or access
control; the upgraded setup is therefore a large improvement from the previous iteration.

The 10 Gb/s switches manage the connections to devices that are involved in the 
transfer of bulk data from the cameras, including the computing nodes, 
storage servers and the CameraPCs of the cameras. These high-throughput switches
enable new, high volume data taking modes thereby increasing the camera capabilities.

The entire onsite network --- encompassing the DAQ machines as well as
the other subsystems, auxiliary devices, personal devices, and external non-H.E.S.S.
devices --- is divided into a few virtual networks
(VLANs) to ensure separation of the network infrastructure into independent parts, for both greater security and stability. This aspect has remained largely unchanged from the
previous iteration. However, in the past, routing tables were used to define permitted
directions of communication between the VLANs; these have been replaced by Access Control Lists
(ACLs) on the switches. 

\subsection{External communication}
\label{sec:external}

A Juniper SRX300 firewall restricts access from the outside to the cluster. 
Previously, the firewall was taken care of via software, using iptables configured 
through FireHOL \cite{firehol}. For the upgrade, it was decided to use a hardware
firewall, as this is a single special purpose device and therefore more robust
to attacks and failures.


\section{Cluster management}

\subsection{Monitoring tools}


The hardware-level monitoring is primarily managed by the Integrated Dell
Remote Access Controller (iDRAC), an implementation of an Intelligent 
Platform Management Interface (IPMI) \cite{IPMI} provided by the manufacturer.
The iDRAC 
provides detailed statuses on the machines' physical components 
such as the power supply, hard disks, cooling devices, etc., including the
history of status changes. It also offers a virtual console and easy
power management options (e.g., powering down, rebooting), all of which are crucial for
a remotely located experiment. The iDRAC replaces the external IPMI devices 
--- previously installed inside each DAQ computer in the old cluster --- and 
web interface.


For monitoring of components and processes related to array operations, we utilize
the TICK Stack \cite{TICK}. This is a set of four pieces of open-source software that can work in tandem:
\begin{enumerate}
    \setlength\itemsep{-0.5em}
    \item Telegraf, for collecting standard system metrics of the monitored machines 
    \item InfluxDB, a time series database for storing the collected metrics
    \item Chronograf, a UI for plotting and viewing the collected metrics
    \item Kapacitor, for metrics processing and alerting
\end{enumerate}
At the end of this flow, if a metric surpasses predetermined limits, the team is 
notified by email as well as Slack messages. Examples of such metrics include high memory 
usage (which could indicate memory leaks caused by bugs during data taking) and 
high storage usage (indicating the need for deleting old data). 
This setup is also used to passively monitor properties such as the data transfer
rate to off-site clusters in Europe, and has been used to identify problems in 
internet connectivity.

\subsection{System provisioning and configuration}

Previously, the system provisioning and configuration was taken care of
with a set of bash scripts. For the upgrade, it was decided to use 
industry standard tools.

The provisioning of the system was performed via a network PXE \cite{PXE} boot,
with the storage servers acting as the PXE-servers. For machine configuration, 
Ansible \cite{Ansible} is used to perform these tasks in a controlled 
and automated manner, by converging the setup to a desired state. It 
is testable, reproducible, uses reusable components and is easily expandable.
It reduces the amount of manual steps and thus the potential for mistakes.
The access to the 
managed machines merely needs ssh and operations can be performed remotely 
and in parallel.

Thanks to this setup, we are able to set up and replace any 
failing cluster components with cold spares, completely from remote.

\section{Integration of the new CT5 camera}

The integration of a new camera for the CT5 telescope took place a few months
after the DAQ upgrade. The installation of a new camera required changes to the
databases, the integration of controllers --- software that acts as the interface
between a subsystem and the DAQ --- for new camera subsystems, and modifications
to the monitoring displays to reflect the properties of the new camera.

The databases contain the configuration information of the various devices. For
the new CT5 camera, tables were added to specify the new camera properties. In addition,
new controller code was written to manage the DAQ-subsystem communications; the new
controllers were then integrated into the definitions of the various run types as
appropriate. 

Part of the DAQ GUI is the \emph{slow control}, a set of processes that produce monitoring 
plots that are updated in realtime. These include camera-related information such as images,
internal temperatures measurements, and voltages, all of which were updated to reflect
the new CT5 camera (e.g., camera shape and drawer configuration). Finally, the
Real-time Analysis was updated to utilize the properties of the new camera.

\section{Evaluation}

Planning for the DAQ upgrade began in mid-2018, about a year before the upgrade took
place. Two weeks were allotted for the actual upgrade; to minimize the loss of
observational time, this overlapped with a moonbreak, during which the 
brightness of the moon precludes the possibility of observations. The goal at the end
of the two weeks was to have a fully functioning array. 
Successful array operations were mostly achieved by this time, 
although minor bug fixes were still required over the following weeks and months.

Given the magnitude of this upgrade --- a full cluster replacement --- this was
only accomplished thanks to thorough preparations. The TestDAQ provided a resource
for testing most aspects of the software beforehand; this proved to be crucial, as 
the libraries, kernel, and
Python versions all had to be updated from the versions on the old cluster, although
the latter was kept at Python2 due to the requirements of some of the legacy
software.
The GlusterFS configuration was set up 
and tested beforehand. The firewall was preconfigured and the ACLs defined, to 
reduce the required amount of setup for the network once onsite. Thanks to these
preparations, the initial setup and installation of the hardware took less than
one day.


The addition of the new CT5 camera required another week of active time, with
a few months of preparation beforehand. Most of the requirements were handled
by the camera team, but the last step in the integration process required action
from the DAQ team.

\begin{figure}
    \centering
    \includegraphics[width=0.65\textwidth]{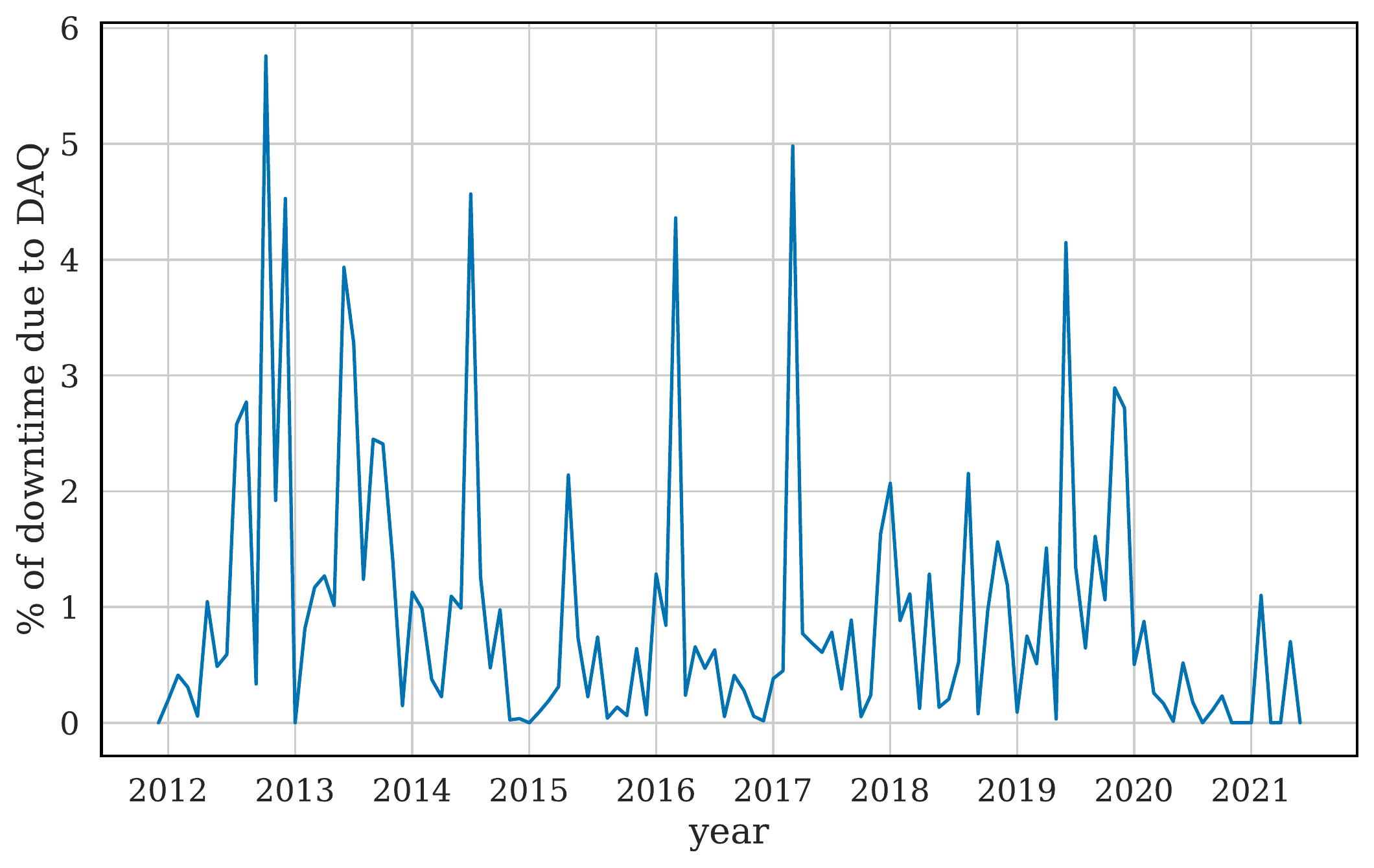}
    \caption{Plotted here is the percentage of the available observation time per month-long shift that is lost due to DAQ-related issues over the last nine years.}
    \label{fig:downtime}
\end{figure}

Thanks to the upgrades, the amount of lost observation time due to issues with
the DAQ is now at an all-time low, as can be
seen from Figure~\ref{fig:downtime}. Before late 2019, a median of 0.64\%
of the available telescope time was lost due to DAQ-related issues, with some 
months reaching nearly 6\%; the median has fallen by a factor of four, 
to just 0.14\% since the start of 2020.

While extensive preparations were undertaken, some difficulties could not 
have been foreseen until the direct interactions with the array hardware. In a
science collaboration, such as H.E.S.S., it is difficult to maintain continuity
of expertise due to the amount of work that is performed by people with 
temporary positions such as students and post docs. This has a very real effect
on not only the extent of possible preparations (which requires identifying who is
in charge of the various pieces of hardware) but also the 
choice of tools for maintaining the
setup; for every new tool, the advantages of the tool must be weighed against its 
learning curve. Thorough documentation, especially with an eye towards future-proofing, 
is therefore crucial for ensuring success.

\end{document}